\documentclass[10pt,twocolumn,amsmath,amssymb,aps,prl,showpacs,longbibliography,superscriptaddress]{revtex4-1}
\usepackage[latin9]{inputenc}
\usepackage{amsmath}
\usepackage{amssymb}
\usepackage{graphicx}
\usepackage{esint}

\makeatletter

\newcommand*\LyXThinSpace{\,\hspace{0pt}}

\usepackage{amsfonts}
\usepackage{graphics}

\makeatother

\begin{document}
\title{Exact spectral properties of Fermi polarons in one-dimensional lattices:
Anomalous Fermi singularities and polaron quasiparticles}
\author{Hui Hu}
\affiliation{Centre for Quantum Technology Theory, Swinburne University of Technology,
Melbourne 3122, Australia}
\author{Jia Wang}
\affiliation{Centre for Quantum Technology Theory, Swinburne University of Technology,
Melbourne 3122, Australia}
\author{Xia-Ji Liu}
\affiliation{Centre for Quantum Technology Theory, Swinburne University of Technology,
Melbourne 3122, Australia}
\date{\today}
\begin{abstract}
We calculate the exact spectral function of a single impurity repulsively
interacting with a bath of fermions in one-dimensional lattices, by
deriving the explicit expression of the form factor for both regular
Bethe states and the irregular spin-flip state and $\eta$-pairing
state, based on the exactly solvable one-dimensional Hubbard model.
While at low impurity momentum $Q\sim0$ the spectral function is
dominated by two power-law Fermi singularities, at large momentum
we observe that the two singularities develop into two-sided distributions
and eventually become anomalous Fermi singularities at the boundary
of the Brillouin zone (i.e., $Q=\pm\pi$), with the power-law tails
extending towards low energy. Near the quarter filling of the Fermi
bath, we also find two broad polaron peaks at large impurity momentum,
collectively contributed by many excited many-body states with non-negligible
form factors. Our exact results of those distinct features in one-dimensional
Fermi polarons, which have no correspondences in two and three dimensions,
could be readily probed in cold-atom laboratories by trapping highly
imbalanced two-component fermionic atoms into one-dimensional optical
lattices.
\end{abstract}
\maketitle
The motion of an impurity interacting with a Fermi bath is a paradigm
in many-body physics and condensed matter physics. It underlies the
fundamental concept of polaron quasiparticle \citep{Alexandrov2010}
and manifests in a variety of intriguing quantum many-body phenomena,
including Anderson orthogonality catastrophe \citep{Anderson1967},
the Fermi edge singularity in the x-ray absorption of metals \citep{Mahan1967,Nozieres1969},
and Nagaoka ferromagnetism \citep{Nagaoka1966}. In the past two decades,
there is a resurgence of interests on this traditional Fermi polaron
problem, due to the rapid advances in ultracold atomic physics \citep{Massignan2014,Schmidt2018,Wang2023AB}.
By tuning the interatomic interactions of a highly spin-population
imbalanced two-component Fermi gas \citep{Bloch2008,Chin2010}, spectral
properties of Fermi polarons in two and three dimensions can now be
routinely measured using radio-frequency spectroscopy \citep{Schirotzek2009,Zhang2012,Zan2019},
Ramsey interferometry \citep{Cetina2016}, Rabi cycle \citep{Scazza2017,Vivanco2024},
and Raman spectroscopy \citep{Ness2020}, with unprecedented accuracy.
At low momentum, the observed spectral features such as the excited
repulsive polaron branch and molecule-hole continuum have been well
explained by variational Chevy ansatz \citep{Chevy2006,Combescot2008,Parish2013,Liu2019},
diagrammatic theories \citep{Combescot2007,Hu2018,Tajima2019,Hu2022,Hu2024},
functional renormalization group \citep{Schmidt2011,vonMilczewski2024},
Monte Carlo simulations \citep{Prokofev2008,Goulko2016,Ramachandran2025}
and exactly solvable models in the heavy polaron limit \citep{Knap2012,Wang2022PRL,Wang2022PRA}.

\begin{figure}
\begin{centering}
\includegraphics[width=0.5\textwidth]{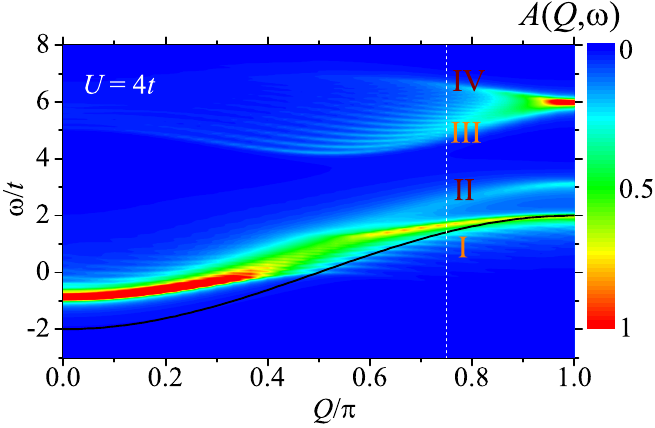}
\par\end{centering}
\caption{\label{fig1_akw2d} The contour plot of the impurity spectral function
$A\left(Q,\omega\right)$, in units of the inverse hopping strength
$t^{-1}$. The black line shows the bare impurity dispersion relation
$\varepsilon_{Q}=-2t\cos Q$. Along the cut at the momentum $Q=0.75\pi$,
as indicated by the vertical dotted line, we may identify four spectral
features I, II, III and IV, to be discussed in detail in Fig. \ref{fig3_excitations}.
Here, we consider a quarter filling $\nu=(N+1)/L=0.5$ with $L=80$
and take an on-site repulsion $U=4t$.}
\end{figure}

In this Letter, by exactly calculating the spectral function based
on the integrable one-dimensional (1D) Hubbard model \citep{Lieb1968,Deguchi2000,Essler2005},
we show that additional distinct characteristics of Fermi polarons,
such as anomalous Fermi singularities (see the features I and III
in Fig. \ref{fig1_akw2d}) in coexistence with polaron quasiparticles
(see the features II and IV), could arise in 1D lattices at large
impurity momentum. These new features are entirely not anticipated.
The 1D Fermi polaron is often considered to be trivial, since Anderson
orthogonality catastrophe survives for a mobile impurity in one-dimension
and therefore Fermi edge singularities are expected to appear in the
impurity spectral function. This idea was partly examined in Ref.
\citep{Castella1993} by calculating the ground-state form factor
of the exactly solvable continuous Gaudin-Yang model \citep{McGuire1966}. 

Here, we generalize the pioneering work \citep{Castella1993} to the
1D Hubbard model and calculate the form factors of all the excited
many-body states, including the spin-flip state and $\eta$-pairing
state \citep{Yang1989} that are not covered by regular Bethe wavefunctions
\citep{Deguchi2000,Essler1992}. This enables us to exactly determine
the impurity spectral function at finite on-site repulsions $U\neq\infty$
for the first time. At low momentum, we observe the conventional Fermi
edge singularities, as already clarified by the calculation of the
ground-state form factor \citep{Castella1993}. With increasing momentum
these Fermi singularities gradually turn into two-sided singularities
and eventually become anomalous with low-energy oriented power-law
tails, upon reaching the boundary of Brillouin zone. Most interestingly,
polaron quasiparticles contributed by a collection of excited many-body
states start to appear, when the filling factor of the Fermi bath
is about quarter.

In addition to providing benchmark results for approximate theories
on Fermi polarons, our exact predictions of new distinct polaron spectral
features could be directly examined in the future cold-atom experiments.
Our technique of determining the form factor is useful to study exact
dynamics of impurity in a lattice Fermi bath such as quantum flutter
\citep{Mathy2012,Knap2014,Gamayun2018,Dolgirev2021,Zhang2024} in
lattices. It may also be generalized to the case of several impurities,
so the effective bath-induced interactions between polaron quasiparticles
and Fermi singularities might be elucidated in the impurity spectral
function. In this case, we may also anticipate to see the emergence
of the spin degree of freedom and the demonstration of the peculiar
spin-charge separation in the spectrum \citep{Deguchi2000,Essler2005,Kohno2010}.

\textbf{\textit{Regular Bethe wavefunctions}}. We start by considering
the well-known Bethe ansatz solution $\left|\Psi_{N+1,Q}\{k_{j},\Lambda\}\right\rangle $
of the 1D Hubbard model with $N$ spin-up fermions and one spin-down
fermion (i.e., the impurity) on a lattice with an even number of sites
$L\geqslant N$ \citep{Lieb1968,Deguchi2000}:
\begin{equation}
\sin k_{j}-\Lambda=u\cot\left(\frac{k_{j}L}{2}\right),\label{eq:BetheAnsatzRootEquation}
\end{equation}
where $u=U/(4t)$ is the dimensionless interaction parameter, $Q=\sum_{j=1}^{N+1}k_{j}$
(mod $2\pi$) is the total momentum, and each real quasi-momentum
$k_{j}$ can be uniquely characterized by an integer $s_{j}\subseteq[-L/2+1,L/2]$
in the region $(s_{j}-1)2\pi/L\leq k_{j}<s_{j}2\pi/L$ \citep{NoteSj,Liu2025}.
In the case of a single impurity, the quasi-momentum $\Lambda$ related
to the impurity position is always real. In contrast, a complex-valued
pair of the quasi-momenta ($k_{1}$ and $k_{2}=k_{1}^{*}$, for clarity)
may arise even for repulsive interaction $u>0$, due to the existence
of two-body bound states \citep{Deguchi2000}. Thus, we classify the
whole regular Bethe wavefunctions with finite quasi-momenta into two
types \citep{Deguchi2000}: the real-$k$ solutions and the $k-\Lambda$
solutions (with a complex-valued pair of quasi-momenta $k_{1}$ and
$k_{2}$).

For the given set of integers $\{s_{j}\}$, Eq. (\ref{eq:BetheAnsatzRootEquation})
can be easily solved for each momentum $k_{j}$, with an adjustable
$\Lambda$ to reproduce the given total momentum $Q$ \citep{Liu2025}.
At a repulsive on-site interaction $u>0$, the ground state has $s_{j}=-(N+1)/2+j$
for an odd $N$ (see Fig. (\ref{fig3_excitations}a)), which might
be visualized as a pseudo Fermi sea. The set of $\{s_{j}\}$ for an
arbitrary excited many-body state can then be uniquely specified by
listing pseudo particle-hole pairs relative to the fully occupied
pseudo Fermi sea. Typically, the energy $E_{N+1}(\{k_{j}\},\Lambda)=-2t\sum_{j=1}^{N+1}\cos k_{j}$
will increase with creating more particle-hole pairs.

Once we solve the quasi-momenta $\{k_{j}\}$ and $\Lambda$, it is
straightforward to write the regular Bethe wavefunction (with a normalization
factor $C_{\Psi}$) into a determinant,
\begin{equation}
\left|\Psi_{N+1,Q}\{k_{j},\Lambda\}\right\rangle =C_{\Psi}\sqrt{\frac{1}{N!L}}e^{iQx_{\downarrow}}\det_{1\leq j,m\leq N}\left[\phi_{j}\left(y_{m}\right)\right],\label{eq:BetheWavefunction}
\end{equation}
where $x_{\downarrow}$ is the position of the impurity and $y_{m}=x_{m}-x_{\downarrow}$
is the coordinate of the $m$-th spin-up fermion relative to the impurity,
and $\phi_{j}(y)\equiv\chi_{j}(y)-\chi_{N+1}(y)$ with the wavefunction
$\chi_{j}(y)=[\Lambda-\sin(k_{j})+iu]e^{ik_{j}y}/(u\sqrt{L})$. This
elegant determinant form of the Bethe wavefunction was first suggested
by Edwards \citep{Edwards1990} and later was used to address the
Fermi edge singularity \citep{Castella1993} and quantum flutter \citep{Mathy2012}
in the Gaudin-Yang model. Here, we adopt a single-particle wavefunction
$\chi_{j}(y)$ that is more suitable to handle the $k-\Lambda$ solutions
with a pair of complex-valued quasi-momenta \citep{Gamayun2015}.

\textbf{\textit{Form factor}}. We are interested in the form factor
or the overlap $F_{N+1}(\{k_{j}\},\Lambda)$ of the Bethe wavefunction
with the product state of the impurity plane-wave (with momentum $Q$)
and a non-interacting Fermi sea, which also takes the form of a Slater
determinant in the first quantization, i.e., $\psi_{Q\downarrow}^{\dagger}\left|\textrm{FS}_{N}\right\rangle =e^{iQx_{\downarrow}}\det_{1\leq j,m\leq N}[e^{iq_{j}y_{m}}/\sqrt{L}]/\sqrt{N!L}$,
where $q_{j}=[-(N+1)/2+j]2\pi/L$. This overlap of two Slater determinants
can be readily obtained, following a well-known identity in quantum
chemistry \citep{Plasser2016},
\begin{equation}
F_{N+1}\left(\{k_{j}\},\Lambda\right)=C_{\Psi}\left|\begin{array}{cccc}
B_{11} & \cdots & B_{1N} & B_{1,N+1}\\
\vdots & \vdots & \vdots & \vdots\\
B_{N1} & \cdots & B_{NN} & B_{N,N+1}\\
1 & \cdots & 1 & 1
\end{array}\right|,\label{eq:FormFactor}
\end{equation}
where the matrix element $B_{jl}\equiv[e^{-iq_{j}}+e^{ik_{l}}]/[L(\sin q_{j}-\sin k_{l})]$.
Using the same identity, we find the normalization factor, $C_{\Psi}=[s_{k\Lambda}u_{1}u_{2}\cdots u_{N+1}(u_{1}^{-1}+u_{2}^{-1}+\cdots+u_{N+1}^{-1})]^{-1/2}$,
where $u_{j}\equiv1+(\sin k_{j}-\Lambda)^{2}/u^{2}+2\cos k_{j}/(Lu)$
and a minus sign $s_{k\Lambda}=-1$ arises for the $k-\Lambda$ solutions.
By introducing the residue $Z_{N+1}(\{k_{j}\},\Lambda)\equiv\left|F_{N+1}(\{k_{j}\},\Lambda)\right|^{2}$,
we obtain the spectral function,
\begin{equation}
A\left(Q,\omega\right)=-\frac{\textrm{Im}}{\pi}\sum_{\{k_{j}\},\Lambda}\frac{Z_{N+1}\left(\{k_{j}\},\Lambda\right)}{\omega-E_{N+1}\left(\{k_{j}\},\Lambda\right)+E_{\textrm{FS},N}+i\delta},
\end{equation}
where $E_{\textrm{FS},N}$ is the energy of the non-interacting Fermi
sea $\left|\textrm{FS}_{N}\right\rangle $ and $\delta\equiv4t/L$
is a broadening factor used to smooth the discrete energy levels at
finite $L$.

\begin{figure*}
\begin{centering}
\includegraphics[width=0.33\textwidth]{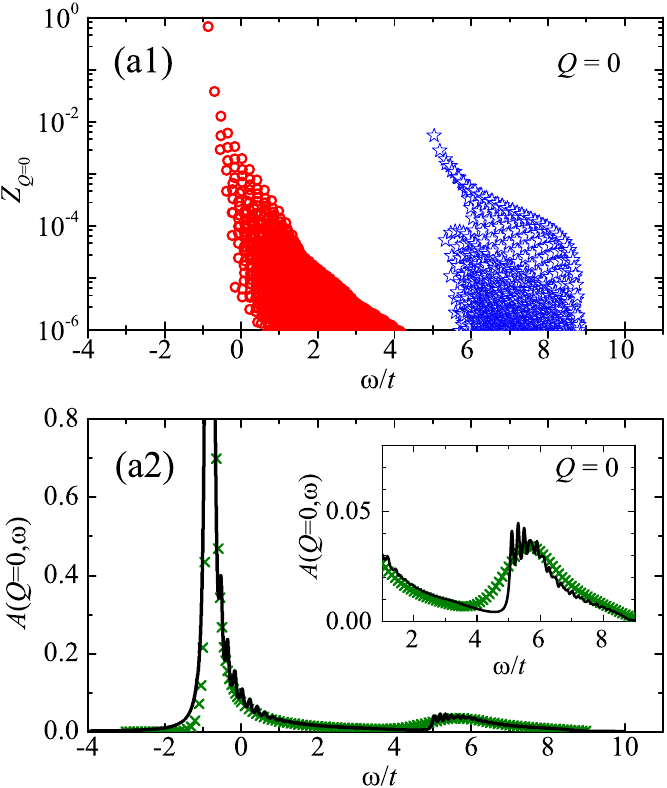}\includegraphics[width=0.33\textwidth]{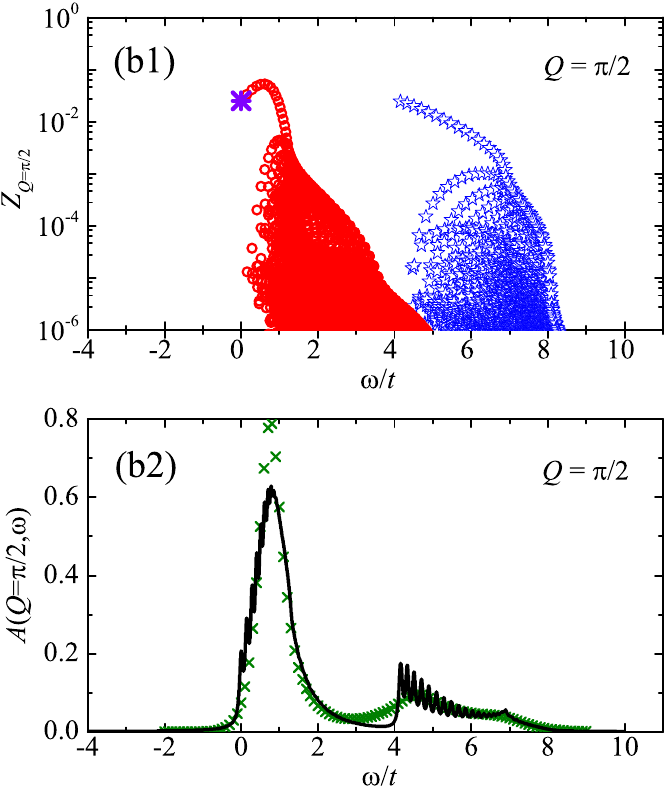}\includegraphics[width=0.33\textwidth]{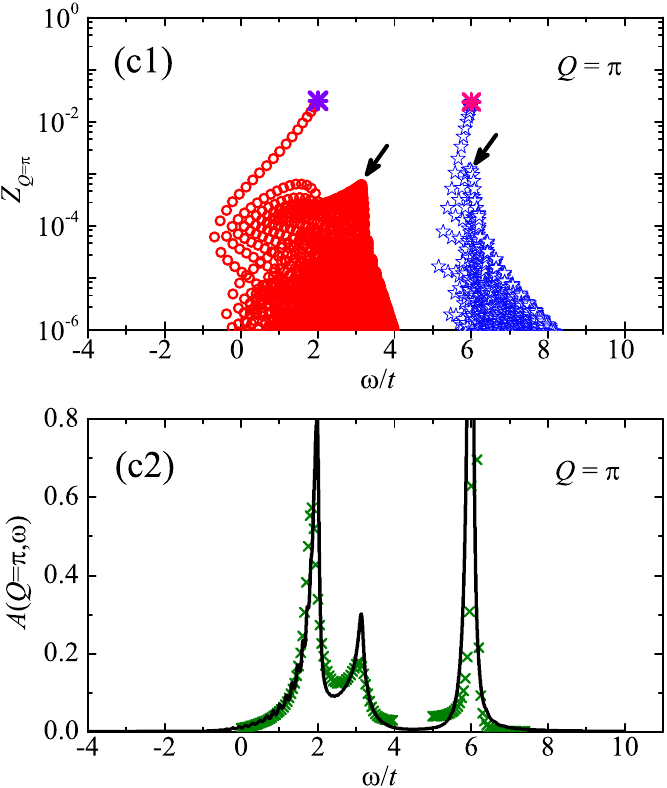}
\par\end{centering}
\caption{\label{fig2_depQ} Residues (a1, b1 and c1) and the impurity spectral
functions (a2, b2 and c2, in units of $t^{-1}$) as a function of
the energy $E_{N+1}-E_{\textrm{FS},N}$ or $\omega$, at different
momentum as indicated. In the upper panel, the residues of the real-$k$
solutions and the $k-\Lambda$ solutions are shown by red circles
and blue stars, respectively. The eight-spoked asterisks show the
residues of the spin-flip state and of the $\eta$-pairing state.
In the lower panel, the green crosses show the spectral functions
predicted by the variational Chevy ansatz with two-particle-hole excitations
\citep{Hu2024} (see Appendix A for more details). The inset in (a2)
highlights the high-energy Fermi singularity.}
\end{figure*}

\textbf{\textit{Irregular states}}. The sum in the above spectral
function should be over all the many-body wavefunctions, including
some irregular states that are not covered by the regular Bethe wavefunctions
\citep{Deguchi2000,Essler1992}. Fortunately, these irregular states
can be directly constructed by acting a spin-flip operator $\zeta^{\dagger}\equiv\sum_{k}\psi_{k\downarrow}^{\dagger}\psi_{k\uparrow}$
and an $\eta$-pairing operator $\eta^{\dagger}\equiv\sum_{k}\psi_{k\downarrow}^{\dagger}\psi_{\pi-k\uparrow}^{\dagger}$
on some non-interacting states \citep{Deguchi2000,Essler1992}. By
inspection, we may explicitly determine the relevant spin-flip state
and $\eta$-pairing state, $\Psi_{\zeta}=C_{\zeta}\zeta^{\dagger}\psi{}_{Q\uparrow}^{\dagger}\left|\textrm{FS}_{N}\right\rangle $
and $\Psi_{\eta}=C_{\eta}\eta^{\dagger}\psi{}_{\pi-Q\uparrow}\left|\textrm{FS}_{N}\right\rangle $,
with energy $E_{\zeta}-E_{\textrm{FS},N}=-2t\cos Q$ and $E_{\eta}-E_{\textrm{FS},N}=-2t\cos Q+U$
and with form factor $F_{\zeta}=C_{\zeta}=1/\sqrt{N+1}$ and $F_{\eta}=C_{\eta}=1/\sqrt{L-N+1}$,
respectively. From the wavefunctions, it is easy to see that the spin-flip
state or $\eta$-pairing state exists only when the momentum $Q$
is out of the Fermi sea (i.e., $Q\notin\{q_{j}\}$) or the momentum
$\pi-Q$ is in the Fermi sea ($\pi-Q\in\{q_{j}\}$). It is worth noting
that when the irregular states exist, we can use the sum rule $\varrho_{s}=F_{\zeta}^{2}+F_{\eta}^{2}+\sum_{\{k_{j}\},\Lambda}Z_{N+1}(\{k_{j}\},\Lambda)=1$
to examine the completeness of the many-body wavefunctions included
in the numerical calculations \citep{Mathy2012}.

\textbf{\textit{Spectral function}}. We calculate the impurity spectral
function for various lattice sizes, filling factors and on-site interactions.
We keep up to three pseudo particle-hole pairs and sum over about
half a million excited states (i.e., for the typical case of $L=80$,
$\nu=(N+1)/L=0.5$ and $U=4t$ unless stated otherwise) with $\varrho_{s}>99.8\%$,
so the omitted states have negligible effect on the convergence of
our numerics. 

In Fig. \ref{fig2_depQ}, we report the residues of all the many-body
states (upper panel) and the related spectral function (lower panel)
at different total momentum \citep{NoteSmallOscillations}. For residues,
the real-$k$ states and the $k-\Lambda$ solutions form two clusters,
roughly separated in energy by the on-site repulsion $U=4t$. Interestingly,
the distributions of the residues in each cluster are qualitatively
similar. At zero momentum $Q=0$, the ground-state residue is dominant
\citep{Castella1993}, followed by a series of states with rapidly
decreasing residues, thus leading to a sharp Fermi edge singularity
at the threshold $\omega_{0}\simeq-t$. In the $k-\Lambda$ cluster,
there is a similar lowest-energy pair state but with a residue $Z\sim0.01$,
giving rise to a much weaker high-energy Fermi singularity at another
threshold $\omega_{1}\sim5t$, as highlighted in the inset of Fig.
\ref{fig2_depQ}(a2). These observations provide a direct confirmation
of the existence of Fermi edge singularities in the spectral function
of 1D Fermi polarons, as suggested in Ref. \citep{Castella1993}.

As the total momentum $Q$ becomes larger than the Fermi wavevector
$k_{F}=N\pi/L$, the residue distributions in the two clusters significantly
change. In particular, the lowest-energy state in each cluster may
no longer have the largest residue. Instead, a branch of states, involving
also the irregular spin-flip state and $\eta$-pairing state, starts
to contribute most significantly, as we shall analyze in detail. At
$Q=\pi/2$ as illustrated in Figs. \ref{fig2_depQ}(b1) and \ref{fig2_depQ}(b2),
this creates a two-sided Fermi singularity in each cluster. As the
momentum $Q$ increases further, the shape of the two-sided Ferm singularity
changes. Upon reaching the boundary of the Brillouin zone at $Q=\pi$,
we find that either the spin-flip state at $\omega=-2t\cos Q=2t$
or the $\eta$-pairing state at $\omega=2t+U$ becomes the state of
the largest residue (see, i.e., the eight-spoked asterisks in Fig.
\ref{fig2_depQ}(c1)). As a consequence, the high-energy side of the
two-sided singularity disappears, leaving an anomalous Fermi singularity
with a power-law tail extending to the low-frequency.

Remarkably, at $Q=\pi$ we observe an additional peak at $\omega\sim3t$,
just above the lower anomalous Fermi singularity at $\omega=2t$,
as shown in Fig. \ref{fig2_depQ}(c2). From the residue distributions,
it is clear that this peak comes from a bundle of many-body states
with a similar residue $Z\sim10^{-3}$, as indicated by the left arrow
in Fig. \ref{fig2_depQ}(c1). Although the residue is small, the number
of the many-body states is sufficiently large to make the peak visible.
This collective nature clearly suggests that the peak is a polaron
quasiparticle, rather than a Fermi singularity. In the $k-\Lambda$
cluster, we also find a similar bundle of many-body states at $\omega=6t$
with $Z\sim10^{-3}$ (see the arrow on the right). However, their
contributions seem to be covered up by the anomalous Fermi singularity
exactly located at the same energy.

\begin{figure}
\begin{centering}
\includegraphics[width=0.45\textwidth]{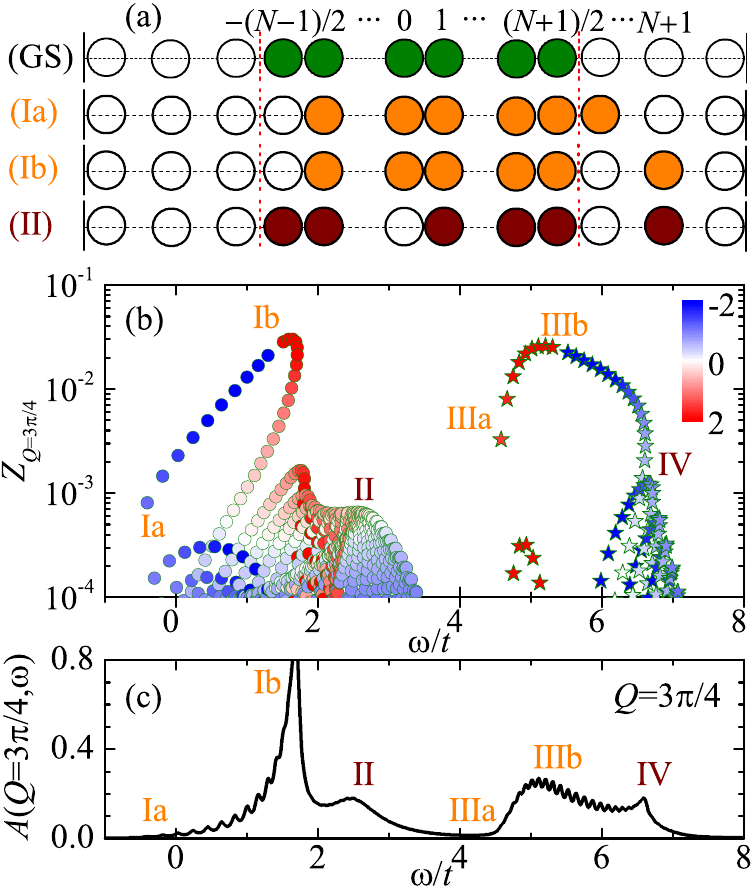}
\par\end{centering}
\caption{\label{fig3_excitations} Residues (b) and the impurity spectral function
(c, in units of $t^{-1}$) at the momentum $Q=3\pi/4$. The color
in the symbols for residues shows the value of the quasi-momentum
$\Lambda$, as indicated by the color-map. The four types of excitations
have been denoted by the Roman numerals I, II, III and IV. The types
Ia, Ib and II are explained diagrammatically in (a), with an illustration
of the ground state (GS) distribution.}
\end{figure}

\textbf{\textit{Nature of singularities and polarons}}. The observation
of the anomalous Fermi singularities at large $Q$, coexisting with
the polaron peaks, is the key result of our Letter. The evolution
of these intriguing features with increasing momentum is displayed
in the two-dimensional contour plot Fig. \ref{fig1_akw2d}. To better
understand how these features arise, we make a cut in the contour
plot at $Q=3\pi/4$ and present the corresponding residues and spectral
function in Fig. \ref{fig3_excitations}. For the residues shown by
symbols in Fig. \ref{fig3_excitations}(b), we assign different colors
to indicate the value of the quasi-momentum $\Lambda$.

We find that the two branches of states that lead to the anomalous
Fermi singularity (I) and the two-sided singularity (III) are well-characterized
by the one pseudo particle-hole excitations (i.e., holon-antiholon
pair \citep{Deguchi2000,Kohno2010}). For example, focusing on the
singularity I, the two states labelled by Ia and Ib in Fig. \ref{fig3_excitations}(b)
are obtained by removing the leftest hole state with $s_{j}=-(N-1)/2$
in the pseudo Fermi sea (GS) and then by occupying the particle state
$s_{j}=(N+1)/2+1$ and $s_{j}=N+1$, respectively, as shown in Fig.
\ref{fig3_excitations}(a). These branches of states are asymmetrically
distributed in $\{s_{j}\}$ compared with the pseudo Fermi sea, and
hence have a significant overlap with the product state $\psi_{Q\downarrow}^{\dagger}\left|\textrm{FS}_{N}\right\rangle $
that is asymmetric itself. We find that such an asymmetry in the distribution
$\{s_{j}\}$ is indirectly evidenced by a large value of the quasi-momentum
$\left|\Lambda\right|$.

In contrast, the polaron peak II is contributed by the many-body states
with nearly zero quasi-momentum $\left|\Lambda\right|\sim0$ (as indicated
by the white color of the symbols), which suggests a symmetric distribution
in $\{s_{j}\}$. As illustrated at the bottom of Fig. \ref{fig3_excitations}(a),
these many-body states are constructed by relocating the hole states
with $s_{j}\sim0$ to the particle states with $s_{j}\sim N+1$, reminiscent
of the bottom-of-band excitations observed in heavy Fermi polarons
(see, i.e., Fig. 6(b) in Ref. \citep{Schmidt2018}). Their residues
are small but the number of such many-body states is large, due to
many different ways of relocation. We note that, the many-body states
forming the polaron II can hardly be understood as the spinon excitations,
which correspond to the change of a quantum number $J$ that characterizes
the quasi-momentum $\Lambda$. However, for multiple impurities, where
a spinon Fermi sea forms, the spinon excitations would show up in
the spectral function near $Q=0$ (or $Q=\pi$) in the lower (or upper)
Hubbard band \citep{Kohno2010}, signaling spin-charge separation.

\begin{figure}
\begin{centering}
\includegraphics[width=0.45\textwidth]{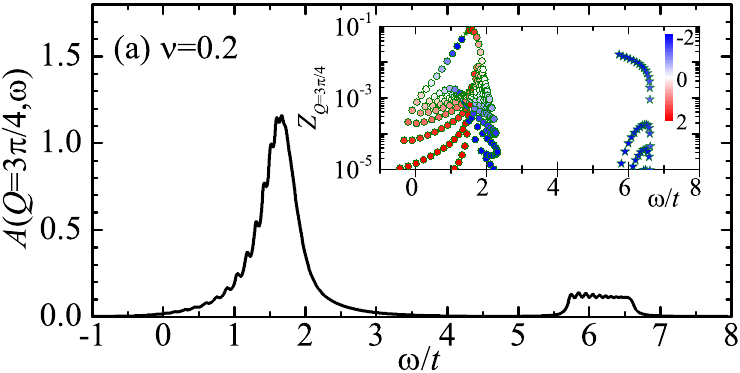}
\par\end{centering}
\begin{centering}
\includegraphics[width=0.45\textwidth]{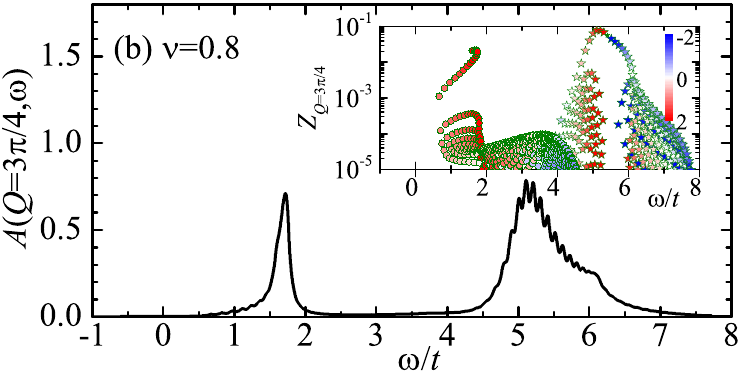}
\par\end{centering}
\caption{\label{fig4_depNu} Impurity spectral functions, in units of $t^{-1}$,
at two filling factors $\nu=0.2$ (a) and $\nu=0.8$ (b) and at the
momentum $Q=3\pi/4$. The insets show the distribution of the residues,
with increasing energy.}
\end{figure}

To further elucidate the existence of singularities and polarons,
we perform a finite-size-scaling analysis of the maximum values $A_{\max}$
for the three peaks in Fig. \ref{fig2_depQ}(c2). By calculating $A_{\max}$
at different length ($L=20$, $40$, $60$ and $80$) with a broadening
factor $\delta=4t/L$, we anticipate a constant $A_{\max}$ for polarons
or a divergent $A_{\max}$ in the form of $\delta^{-\alpha}$ ($0<\alpha<1$)
for singularities as $\delta\rightarrow0$. We find $\alpha=0.50$
and $\alpha=0.89$ for the first and third peaks, indicating their
nature as Fermi singularities. Instead, we obtain a much smaller exponent
$\alpha=0.31$ for the second peak at $\omega\sim3t$, which is more
consistent with the interpretation of a polaron quasiparticle. More
details are given in Appendix B.

We finally note that, the co-existence of singularities and polarons
at $Q\sim\pi$ take place only near the quarter filling factor $\nu=1/2$.
This might be related to the backward scattering between the two Fermi
points, which becomes favourable when the transferred momentum $2k_{F}\sim\pi\sim Q$.
In the dilute limit (see Fig. \ref{fig4_depNu}(a) for $\nu=0.2$),
the separated clustering of the many-body states for the polaron quasiparticle,
as indicated by II in Fig. \ref{fig3_excitations}(b), does not occur.
Near the half filling (Fig. \ref{fig4_depNu}(b)), the clustering
does happen. However, the residues of the many-body states are too
small to form a visible polaron peak.

\textbf{\textit{Conclusions}}. In summary, we have calculated the
exact spectral function of one-dimensional Fermi polarons in lattices,
by using the celebrated Bethe wavefunctions \citep{Deguchi2000}.
We have found the coexistence of anomalous Fermi singularity and polaron
quasiparticles at large momentum near the quarter filling. This interesting
feature could be experimentally observed with ultracold atoms using
Ramsey interferometry \citep{Dolgirev2021}, in which the impurity
can be first accelerated to acquire momentum $Q$ \citep{Meinert2017}
and the consequently measured Ramsey overlap function $S(t)$ gives
rise to the spectral function $A(Q,\omega)$ after a Fourier transformation
\citep{Schmidt2018,Knap2012}. Further extension of our work with
multiple impurities would be useful to explore polaron-polaron interactions
and the emergence of spin-charge separation in one dimension \citep{Essler2005,Kohno2010}. 
\begin{acknowledgments}
\textbf{\textit{Acknowledgments}}. This research was supported by
the Australian Research Council's (ARC) Discovery Program, Grants
Nos. DP240101590 (H.H.), FT230100229 (J.W.), and DP240100248 (X.-J.L.).
\end{acknowledgments}

\appendix

\section{Spectral function from Chevy ansatz approach}

\begin{figure}
\begin{centering}
\includegraphics[width=0.45\textwidth]{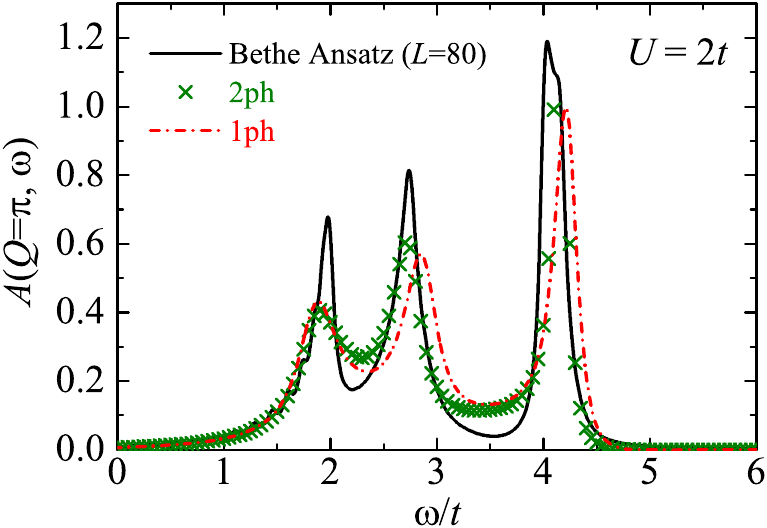}
\par\end{centering}
\caption{\label{fig5_Chevy2ph} Impurity spectral function $A(Q=\pi,\omega)$
predicted by using variational Chevy ansatz, with one-particle-hole
excitations (red dot-dashed line) and with two-particle-hole excitations
(green crosses) \citep{Hu2024} in the thermodynamic limit, with a
broadening factor $\delta$ extrapolated to zero. Here, we take the
on-site repulsion $U=2t$. For the Bethe ansatz calculation, the number
of spin-up atoms is $N=39$ and the number of sites is $L=80$.}
\end{figure}

In previous works, the polaron spectral function has been calculated
by using various approximate approaches, including variational Chevy
ansatz and many-body diagrammatic theory \citep{Hu2024}. In the lower
panel of Fig. \ref{fig2_depQ}, we have compared the exact predictions
from Bethe ansatz (solid lines) with the results from variational
Chevy ansatz with the inclusion of two-particle-hole excitations (green
crosses). In Fig. \ref{fig5_Chevy2ph}, we present an additional comparison
at the on-site repulsion $U=2t$, including the variational results
with both one-particle-hole excitations (see the red dot-dashed line)
and two-particle-hole excitations. 

Using the exact Bethe ansatz results as a benchmark, we find that
the inclusion of two-particle-hole excitations quantitatively improves
the predictive power of the variational Chevy ansatz. In particular,
the polaron peak in the middle seems to be well-reproduced by the
Chevy ansatz approach with two-particle-hole excitations.

\begin{figure}
\begin{centering}
\includegraphics[width=0.45\textwidth]{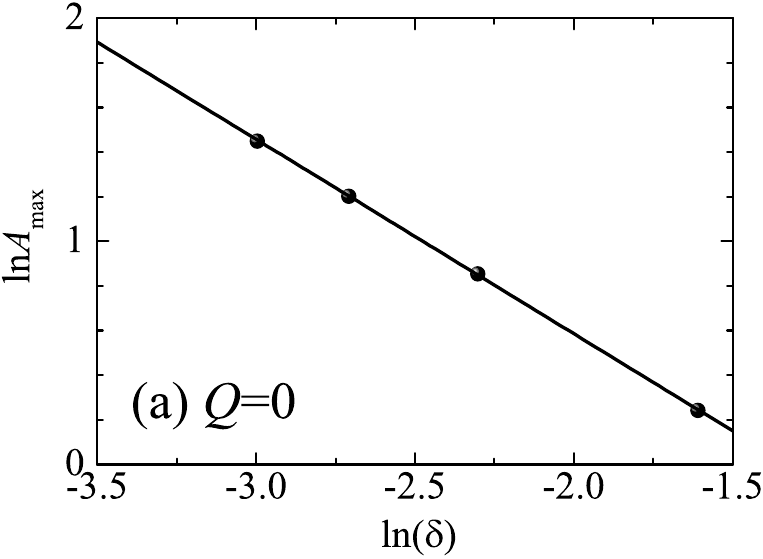}
\par\end{centering}
\begin{centering}
\includegraphics[width=0.45\textwidth]{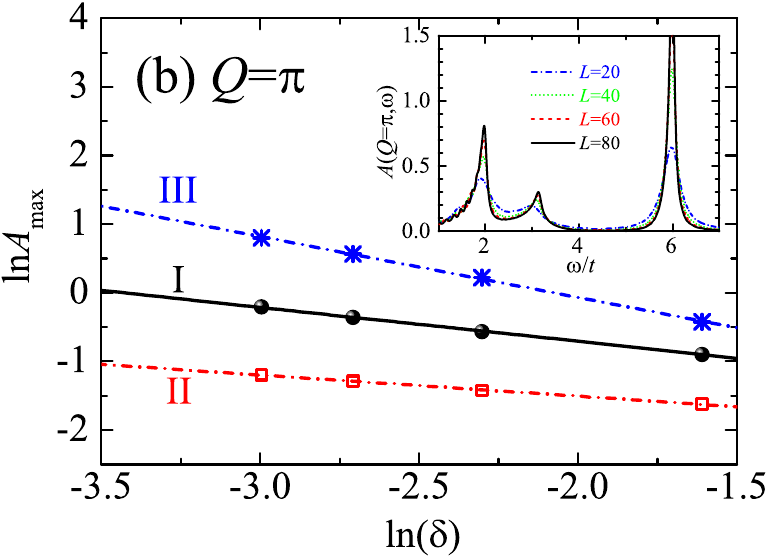}
\par\end{centering}
\caption{\label{fig6_scaling} $\ln A_{\max}$ as a function of $\ln\delta$,
for the $Q=0$ singularity (a) and for the singularities (I and III)
and the polaron peak (II) at $Q=\pi$ (b). The lines are the linear
fits, $\ln A_{\max}=-\alpha\ln\delta+\textrm{const}$, to the data
points. The inset in (b) reports the spectral function at different
number of sites $L$. Here, we consider a quarter filling $\nu=(N+1)/L=0.5$
and an on-site repulsion $U=4t$.}
\end{figure}

\section{Finite-size-scaling analysis}

In the main text, we have stated that the maximum value $A_{\max}$
of a Fermi singularity scales as, $A_{\max}\propto\delta^{-\alpha}$,
where $\delta=4t/L$ is the broadening factor and the exponent $\alpha$
satisfies $0<\alpha<1$. For the polaron peak, we instead anticipate
a constant $A_{\max}$, independent of the number of sites $L$ (i.e.,
$\alpha\sim0$).

To check this scaling law, in Fig. \ref{fig6_scaling}(a) we plot
$\ln A_{\max}$ as a function of $\ln\delta$ for the Fermi-edge singularity
at $Q=0$ and at $U=4t$ and half filling $\nu=0.5$. In this case,
the exponent $\alpha$ is analytically given by \citep{Castella1993},
\begin{equation}
\alpha=1-2\left(\frac{\delta'_{F}}{\pi}\right)^{2},
\end{equation}
where the phase shift $\delta_{F}^{'}$ is related to the on-site
repulsion by, $\delta_{F}^{'}=-\arctan(\pi Un_{F}/2)$, where $n_{F}=1/(2\pi t\sin k_{F})$
is the density of state at the Fermi point and $k_{F}=\nu\pi$ is
the Fermi wavevector. By substituting $U=4t$ into the the expressions,
we find that the phase shift $\delta_{F}^{'}=-\pi/4$ and hence the
exponent $\alpha=7/8=0.875$. Numerically, by linearly fitting the
data in Fig. \ref{fig6_scaling}(a), we obtain from the slope $\alpha_{\textrm{fit}}=0.871$,
which is very close to the expected value of $\alpha=0.875$.

The linear fits for the anomalous Fermi singularities (I and III)
and the polaron peak (II) at $Q=\pi$ are reported in Fig. \ref{fig6_scaling}(b).
The exponent for the polaron peak (II), $\alpha_{\textrm{II}}=0.31$,
is small but nonzero. This is probably due to the fact that the polaron
peak (II) is not completely separated from the Fermi singularity (I)
due to a finite broadening factor $\delta$. The finite overlap then
may bring a weak $L$-dependence to the polaron peak value $A_{\max}$.

\begin{figure}
\begin{centering}
\includegraphics[width=0.45\textwidth]{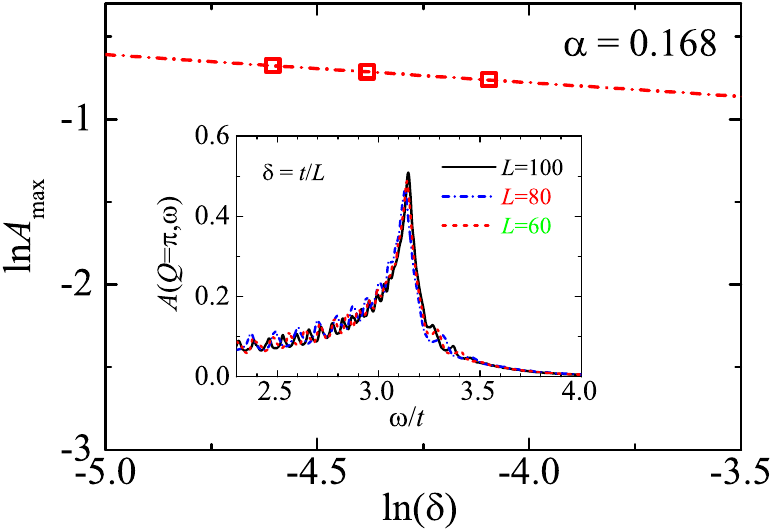}
\par\end{centering}
\caption{\label{fig7_scailingII} $\ln A_{\max}$ as a function of $\ln\delta$,
for the the polaron peak (II) at $Q=\pi$, where we take a smaller
broadening factor $\delta=t/L$. The linear fit (i.e., $\ln A_{\max}=-\alpha\ln\delta+\textrm{const}$),
as given by the red dot-dashed line, leads to an exponent $\alpha_{\textrm{II}}\simeq0.17$.
The inset reports the spectral function at different number of sites
$L$ for the polaron peak (II). The other parameters are the same
as in Fig. \ref{fig6_scaling}.}
\end{figure}

To reduce the overlap, we take a smaller broadening factor $\delta=t/L$
with $L=60$, $80$ and $100$ for the collective peak (II), as shown
in Fig. \ref{fig7_scailingII}. The numerical fit of the data $\ln A_{\max}$
leads to an exponent $\alpha_{\textrm{II}}=0.17$. As anticipated,
this exponent is indeed smaller than the value of $0.31$, obtain
with $\delta=4t/L$. It is reasonable to believe that, by taking the
limit $\delta\rightarrow0$ with more numerical efforts, we might
eventually confirm $\alpha_{\textrm{II}}=0$ for the polaron peak
(II). 

\section{Attractive on-site interaction}

\begin{figure}
\begin{centering}
\includegraphics[width=0.45\textwidth]{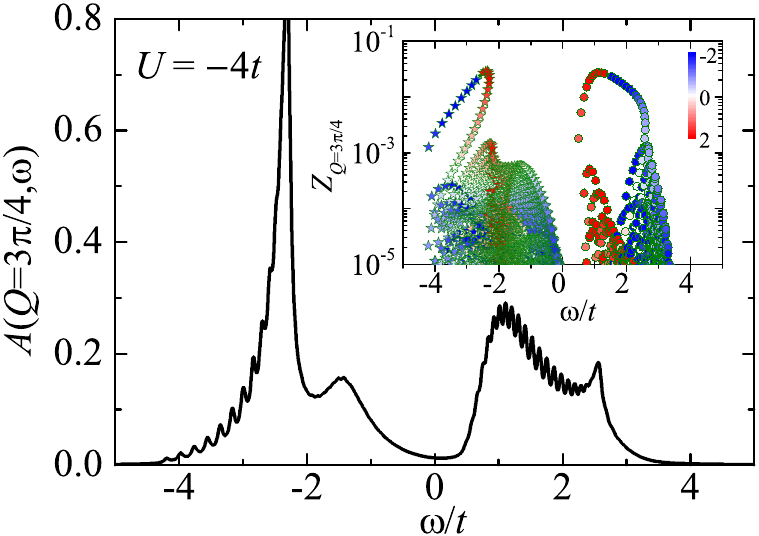}
\par\end{centering}
\caption{\label{fig8_U4tm} Impurity spectral function $A(Q=3\pi/4,\omega)$
at an attractive on-site interaction $U=-4t$. As in Fig. \ref{fig3_excitations}(c),
we consider a quarter filling $\nu=(N+1)/L=0.5$ with $L=80$.}
\end{figure}

Here, we also examine the case of an attractive on-site interaction.
In Fig. \ref{fig8_U4tm}, we show the impurity spectral function $A(Q=3\pi/4,\omega)$
at $U=-4t$, keeping other parameters the same as in Fig. \ref{fig3_excitations}(c).
Interestingly, we find the qualitative same spectral function as in
the case of on-site repulsion $U=4t$, in spite of a constant energy
shift around $4t$. Therefore, the coexistence of the anomalous Fermi
singularities and polaron quasiparticles seems to be a robust and
universal feature of Fermi polaron in 1D lattices, regardless of the
sign of on-site interactions.

This similarity can be easily understood by performing a particle-hole
transformation for the Fermi bath (i.e., spin-up fermions), $\psi_{i\downarrow}^{\dagger}\rightarrow(-1)^{i}\psi_{i\downarrow}$,
under which the single-particle kinetic Hamiltonian is invariant and
up to a constant energy shift the interaction Hamiltonian acquires
a minus sign, i.e., $U\rightarrow-U$. The filling factor of the Fermi
bath changes as, $\nu\rightarrow1-\nu$. Therefore, the polaron problem
at the repulsion ($+U$) at the filling factor $\nu$ is identical
to the case of an attractive interaction ($-U$) at the filling factor
$1-\nu$, if we neglect the unimportant energy shift.


\begin{thebibliography}{99}
\bibitem{NoteSmallOscillations}There are small fast oscillations
in the spectral function, due to the discrete eigenstates at finite
number of lattice sites $L$, which can not completely smoothed out
by the finite broadening factor $\delta=4t/L$. These fast oscillations
are not important and should disappear for sufficiently large $L$.
\end{thebibliography}
\end{document}